\documentclass[preprint,preprintnumbers,amsmath,amssymb]{revtex4}

\usepackage{graphicx}
\usepackage{bm}
\usepackage{feynmp}

\begin{document}

\preprint{USTC-ICTS-08-13}

\title{Consistency Relations for Non-Gaussianity}

\author{Miao Li}

\email{miaoli1@ustc.edu.cn}

\author{Yi Wang}

\email{wangyi@itp.ac.cn}

\affiliation{%
Interdisciplinary Center of Theoretical Studies, USTC, Hefei, Anhui
230026, P.R.China\\ Institute of Theoretical Physics, CAS, Beijing
100080, P.R.China
}%

\date{\today}

\begin{abstract}
We investigate consistency relations for non-Gaussianity. We provide
a model-independent dynamical proof for the consistency relation of
3-point correlation functions from the Hamiltonian and field
redefinition. This relation can be applied to single field
inflation, multi-field inflation and the curvaton scenario. This
relation can also be generalized to $n$-point correlation functions
up to arbitrary order in perturbation theory and with arbitrary
number of loops.
\end{abstract}

\maketitle

\section{Introduction}

The CMB non-Gaussianity \cite{fnl} has been extensively investigated
in the recent years. The 3-point correlation of the curvature
perturbation for a large variety of inflation models has been
calculated, including the standard single field inflation
\cite{NG,NG1}, K-inflation \cite{K}, DBI inflation
\cite{DBI,Chen:2006nt}, multi-field inflation
\cite{multiple,Seery:2005gb,Arroja:2008yy}, the curvaton scenario
\cite{Lyth:2001nq} and the ekpyrotic scenario \cite{ekpyrotic}. The
4-point correlation has been calculated in
\cite{Seery:2006vu,trispectrum}.

Maldacena found that there is a consistency relation for the 3-point
correlation function of single field inflation \cite{NG}. The
consistency relation states that in the limit that one of the three
momenta goes to zero, the 3-point correlation function should be
proportional to the spectral index times the square of the power
spectrum. This consistency relation has been discussed in more
detail in \cite{consistency}. The generalization of this relation to
the 4-point correlation is discussed in \cite{Seery:2006vu}.

The consistency relation is derived in two ways in the literature.
One is the original back-reaction argument by Maldacena, another is
to check the relation in explicit models. In this paper, we shall
derive the consistency relation from the Hamiltonian. The proof is
model independent, and follows from dynamics. The proof also applies
for a general $n$-point correlation function up to arbitrary order
in perturbation theory and arbitrary number of loops. We shall also
derive a consistency relation from a local field redefinition.

The consistency relation we derive can be used not only in single
field inflation, but also in multi-field inflation as well as in the
curvaton scenario. The consistency relation in multi-field models is
also discussed in \cite{Allen:2005ye}.

The paper is organized as follows. In Section II, we discuss the
consistency relation for 3-point correlation functions. We first
setup the basic notations and equations. Then we derive the
consistency relation from the action and local field redefinition.
After that, we apply our relation to single field inflation,
multi-field inflation, and the curvaton scenario. In Section III, we
generalize the derivation of the consistency relation to $n$-point
correlation functions. We conclude in Section IV.

\section{Consistency Relation for the 3-point Function}

\subsection{The Basic Setup}
We use $\phi_a$ to collectively denote fields, derivative of the
fields, and the conjugate momenta. We assume the fields are real. A
complex field can be decomposed into a doublet of real fields
without losing generality. These fields can be expanded around the
homogeneous background,
\begin{equation}
  \phi_a({\bf x},t)=\bar\phi_a(t)+\delta\phi_a({\bf x},t)~,\qquad a=1,2,\ldots, n~.
\end{equation}

The Hamiltonian can be perturbatively expanded as
\begin{equation}
  H[\phi({\bf
  x},t)]=\sum_{N=0}^{\infty}\frac{1}{N!}
  \left(\sum_{a=1}^{n}\delta\phi_a({\bf x},t)\partial_a\right)^N H[\bar\phi(t)]\equiv \sum_{N=0}^{\infty}H_N[\delta\phi({\bf
  x},t),t]~,
\end{equation}
where $\partial_a$ denotes $\frac{\delta}{\delta
  \bar\phi_a(t)}$, which is performed on the Hamiltonian.
 The quantity $\delta\phi_a({\bf x},t)$ is canonically
quantized according to the second order Hamiltonian
$H_2[\delta\phi({\bf x},t),t]$. Terms $H_N$ $(N\geq 3)$ are treated
as interactions,
\begin{equation}
  H_{\rm int}[\delta\phi({\bf x},t),t]\equiv \sum_{N=3}^{\infty}H_N[\delta\phi({\bf
x},t),t]~.
\end{equation}

An expectation value can be calculated using
\begin{align}\label{pert}
  \langle Q(t) \rangle &=\sum_{N=0}^{\infty}i^N
  \int_{-\infty}^{t}dt_{N}\int_{-\infty}^{t_N}dt_{N-1}\cdots\int_{-\infty}^{t_2}dt_{1}
  \nonumber\\&
  \times\left\langle \big[H_{\rm int}(t_1),\big[H_{\rm int}(t_2),\cdots
  \big[H_{\rm int}(t_N),Q^I(t)\big]\cdots\big]\big] \right\rangle~,
\end{align}
where $Q(t)$ is some product of the field operators $\delta\phi_a$,
and $Q^I(t)$ is $Q(t)$ in the interaction picture. The $N=0$ term in
the RHS should be understood to be just $\langle Q^I(t)\rangle$. We
shall omit the superscript $I$ in the following sections for
simplicity.

In the following discussion, we will drop the disconnected part of
the correlation functions. All correlation functions automatically
denote the connected part.

\subsection{Consistency Relation from the Hamiltonian}
The 3-point correlation function can be calculated using
\begin{equation}
  \langle\delta\phi_a({\bf k}_1,t)\delta\phi_b({\bf k}_2,t)\delta\phi_c({\bf
  k}_3,t)\rangle=-i\int_{-\infty}^t dt'
  \left\langle\big[
\delta\phi_a({\bf k}_1,t)\delta\phi_b({\bf k}_2,t)\delta\phi_c({\bf
k}_3,t), H_3(t')
  \big]  \right\rangle~.
\end{equation}

In this paper, we normalize the fields in a box, and use the
convention
\begin{equation}
  \delta\phi_a({\bf x},t)=\frac{1}{\sqrt V}\sum_{\bf k} \delta\phi_a({\bf
  k},t) e^{i {\bf k}\cdot {\bf x}} ~.
\end{equation}
Here $V$ is the space volume. More details about our convention can
be found in the appendix.

The Fourier modes of the field can be written in terms of the
creation and annihilation operators as
\begin{equation}
  \delta\phi_a({\bf k},t)=\delta\phi_a^{\rm (cl)}({\bf k},t) a_{\bf
  k}+\delta\phi_a^{{\rm (cl)}*}({\bf -k},t) a^{\dagger}_{\bf -
  k}~,
\end{equation}
where $\delta\phi_a^{\rm (cl)}({\bf k},t)$ is a $c$-number function
and satisfies the classical equation of motion.

We use $k_i$ to denote $|{\bf k}_i|$. Consider the limit $k_1 \ll
\min (k_2, k_3, aH)$, (we shall refer to this limit as
$k_1\rightarrow 0$ limit for simplicity). In the correlation
function, $\delta\phi_a({\bf k}_1,t)$ can be either to the left or
to the right of the Hamiltonian $H$, corresponding to the term
$\langle \delta\phi^3 H\rangle$ and $\langle H\delta\phi^3\rangle$
respectively. In the $\langle \delta\phi^3 H\rangle$ case, keeping
in mind that the annihilation operator in $\delta\phi_a({\bf
k}_1,t)$ must be contracted with one of the $\delta\phi_d$ in $H_3$,
we have

\begin{align}\label{H3}
  H_3&=\frac{1}{3!}\left(\sum_d\delta\phi_d({\bf
x},t)\partial_d\right)^3 H
\rightarrow\frac{1}{2}\left(\sum_d\delta\phi_d({\bf
x},t)\partial_d\right)^2 \sum_e\left[
\overline{\delta\phi}_e\partial_e \right]H \nonumber\\&
=\frac{1}{2}\left(\sum_d\delta\phi_d({\bf x},t)\partial_d\right)^2
\big(H(\bar\phi+\overline{\delta\phi})-H(\bar\phi)\big)=H_2(\bar\phi+\overline{\delta\phi})-H_2(\bar\phi)~.
\end{align}
where ``$\rightarrow$'' indicates that we have made the contraction
with $\delta\phi_a({\bf k}_1,t)$, and have taken the $k_1\rightarrow
0$ limit. After doing the contraction, the field is replaced by its
classical part, and this replacement also applies for
$\delta\phi_a({\bf k}_1,t)$. $\delta(a,e)$ is defined as,
$\delta(a,e)=1$ if $\delta\phi_a$ and $\delta\phi_e$ are the same
fundamental field, or the derivative or the conjugate momentum of
it, otherwise, $\delta(a,e)=0$.  And
$\overline{\delta\phi}_e\equiv\delta(a,e)V^{-1}\int_V\delta\phi_e^{{\rm
(cl)}*}({\bf x},t)dV$. At last, we have neglected the high order
terms of $\overline{\delta\phi}_e$, because we are considering in
this section the leading order perturbation. In the next section, we
shall show that the consistency relation is respected order by order
in the perturbation theory.

With the help of Eq. \eqref{H3}, the term
$\langle\delta\phi^3H\rangle$ can be written as
\begin{equation}
\langle\delta\phi_{a}({\bf k}_1,t)\delta\phi_{b}({\bf
k}_2,t)\delta\phi_{c}({\bf
k}_3,t)H_3(\bar\phi)\rangle\xrightarrow{k_1\rightarrow
0}\delta\phi_{a}^{\rm (cl)}({\bf
k}_1,t)\left\langle\delta\phi_{b}({\bf k}_2,t)\delta\phi_{c}({\bf
k}_3,t)\big(H_2(\bar\phi)-H_2(\bar\phi-\overline{\delta\phi})\big)\right\rangle~.
\end{equation}
The other term $\langle H\delta\phi^3\rangle$ can be written
similarly. One difference is that the terms $\delta\phi_{a}^{\rm
(cl)}({\bf k}_1,t)$ and the corresponding terms in the Hamiltonian
should be replaced by their complex conjugates. However, when
$k_1\ll aH$, this difference can be neglected. It is because after
horizon crossing, the perturbations become classical, and the mode
functions can be made real by a time-independent phase rotation
\cite{Polarski:1995jg}. So we have $\delta(a,e)\delta\phi_{a}^{\rm
(cl)}({\bf k}_1,\tau)\delta\phi_{e}^{\rm (cl)}(-{\bf
k}_1,\tau')^*=\delta(a,e)\delta\phi_{a}^{\rm (cl)}(-{\bf
k}_1,\tau)^*\delta\phi_{e}^{\rm (cl)}({\bf k}_1,\tau')$.
\footnote{Note that $\tau'$ can run from $-\infty$ to $0$.
$k_1\tau'$ is not always small along the $\int d\tau'$ integration.
However, given a lower cutoff on $\tau'$ (which is due to the choice
of the interacting vacuum), the integral uniformly converges when we
take $k_1\rightarrow 0$ , so we can interchange the $k_1\rightarrow
0$ limit and the integration to obtain
$\delta(a,e)\delta\phi_{a}^{\rm (cl)}({\bf
k}_1,\tau)\delta\phi_{e}^{\rm (cl)}(-{\bf
k}_1,\tau')^*=\delta(a,e)\delta\phi_{a}^{\rm (cl)}(-{\bf
k}_1,\tau)^*\delta\phi_{e}^{\rm (cl)}({\bf k}_1,\tau')$. This can be
verified in models with either standard or generalized kinetic
terms.} This condition can also be checked explicitly in inflation
models with either standard or generalize kinetic terms.

Note that the condition $k_1 \ll aH$ is essential in proving the
consistency relation. If this condition is not satisfied, we can
check explicitly that the consistency relation is not satisfied even
in the simplest single field inflation model with standard kinetic
terms.

Note that the fields in the interaction picture evolves according to
the free Hamiltonian $H_2(\bar\phi)$, not
$H_2(\bar\phi+\overline{\delta\phi})$. The difference between these
two should also be treated as the interaction Hamiltonian. So in the
$k_1\rightarrow 0$ limit we have
\begin{align}
& \langle\delta\phi_a({\bf k}_1,t)\delta\phi_b({\bf
k}_2,t)\delta\phi_c({\bf
  k}_3,t)\rangle
\nonumber\\
  =&-i \delta\phi_a^{\rm (cl)}({\bf k}_1,t)\int_{-\infty}^t dt'\left\langle
  \big[\delta\phi_b({\bf k}_2,t)\delta\phi_c({\bf
k}_3,t), \big(H_2(\bar\phi+\overline{\delta\phi})-H_2(\bar\phi)\big)
\big]\right\rangle \nonumber\\ =&\delta\phi_a^{(\rm cl)}({\bf
k}_1,t)\Big(\langle\delta\phi_b({\bf k}_2,t)\delta\phi_c({\bf
  k}_3,t)\rangle\Big|_{\overline{\delta\phi}}-\langle\delta\phi_b({\bf
k}_2,t)\delta\phi_c({\bf
  k}_3,t)\rangle\Big|_{\overline{\delta\phi}=0}\Big)~,
\end{align}
where in the last line, one should note that for the 2-point
correlation function, the $N=0$ term in Eq. \eqref{pert} is also
present. This equation can be rewritten as
\begin{equation}\label{3point1}
\langle\delta\phi_a({\bf k}_1,t)\delta\phi_b({\bf
k}_2,t)\delta\phi_c({\bf
  k}_3,t)\rangle \xrightarrow{k_1\rightarrow 0}
  \delta\phi_a^{\rm (cl)}({\bf
k}_1,t)\sum_e\overline{\delta\phi}_e\partial_e\langle\delta\phi_b({\bf
k}_2,t)\delta\phi_c({\bf
  k}_3,t)\rangle~.
\end{equation}
The classical field and the quantum expectation can be related as
\begin{equation}
  \delta\phi_a^{\rm (cl)}({\bf k}_1,t)\overline{\delta\phi}_e=\frac{\delta(a,e)}{\sqrt V}\delta\phi_a^{\rm (cl)}({\bf k}_1,t)\delta\phi_e^{{\rm (cl)}*}({\bf
  k}_1,t)=\frac{\delta(a,e)}{V^{3/2}}\langle\delta\phi_a({\bf k}_1,t)\delta\phi_e^*({\bf
  k}_1,t)\rangle~.
\end{equation}
So Eq. \eqref{3point1} can be recast as
\begin{align}\label{3point}
&\langle\delta\phi_a({\bf k}_1,t)\delta\phi_b({\bf
k}_2,t)\delta\phi_c({\bf
  k}_3,t)\rangle \xrightarrow{k_1\rightarrow 0}
\nonumber\\
&  \sum_e \frac{\delta(a,e)}{V}
  \langle\delta\phi_a({\bf k}_1,t)\delta\phi_e^*({\bf
  k}_1,t)\rangle
  \frac{\partial}{\partial \delta\phi_e^{{\rm (cl)}*}({\bf
  k}_1,t)}\langle\delta\phi_b({\bf
k}_2,t)\delta\phi_c({\bf
  k}_3,t)\rangle~.
\end{align}

One should note that in the above proof, we have used the condition
that $\delta\phi_a$ is a perturbation of the background field
$\phi_a$. There are usually two such variables, $\zeta$, as the
perturbation of the logarithm of the scale factor in the uniform
density slice, and $Q_a$, as the perturbation of the inflaton fields
in the flat slice. The resulting Eq. \eqref{3point} can not be
directly used to composite perturbation variables beyond this
condition. The consistency relation for induced variables can be
derived from a field redefinition, discussed in the next subsection.
For single field inflation, the consistency relation for $\zeta$ and
the consistency relation for $Q$ can also be related by such a field
redefinition (see Appendix B).

\subsection{Consistency Relation from Local Field Redefinition}

Sometimes, the field we use in the Hamiltonian and the field we use
in the correlation function are up to a local non-linear
redefinition of the fields. In this case, we need to consider the
contribution to the consistency relation from the field
redefinition. Sometimes, the non-Gaussianity is dominated by this
redefinition, such as in the curvaton scenario. In this subsection,
we shall derive the consistency relation from the field
redefinition.

Firstly, let us consider the ansatz of local non-Gaussianity. This
case can be directly generalized to a more general local field
redefinition. In the local ansatz, the field redefinition takes the
form
\begin{equation}\label{localNG}
  \zeta({\bf x},t)=\zeta_g({\bf x},t)+\frac{3}{5}f_{\rm NL} \zeta_g({\bf
  x},t)^2~,
\end{equation}
where $\zeta_g({\bf x},t)$ is the Gaussian part of $\zeta({\bf
x},t)$. In the Fourier space, Eq. \eqref{localNG} becomes
\begin{equation}\label{3g}
  \zeta_{{\bf k}}=\zeta_{g{\bf k}}+\frac{3}{5}f_{\rm NL}\frac{1}{V}\sum_{k'}\zeta_{g{\bf k'}}\zeta_{g{\bf
  k-k'}}~.
\end{equation}
In the $k_1\rightarrow 0$ limit, the 3-point function takes the form
\begin{align}\label{3r}
   \langle\zeta_{{\bf k}_1}\zeta_{{\bf k}_2}\zeta_{{\bf
  k}_3}\rangle&\xrightarrow{k_1\rightarrow 0}
  \frac{3}{5}f_{\rm NL} \frac{1}{V}\sum_{{\bf k}'}
  \langle\zeta_{g{\bf k}_1}\zeta_{g{\bf k}'}\zeta_{g{\bf k}_2-{\bf k}'}\zeta_{g{\bf
  k}_3}\rangle+({\bf k}_2\Leftrightarrow {\bf k}_3)\nonumber\\&=
  \frac{12}{5}f_{\rm NL}\frac{1}{V} \langle\zeta_{g{\bf k}_1}\zeta_{g{\bf k}_1}^*\rangle\langle\zeta_{g{\bf k}_2}\zeta_{g{\bf
  k}_3}\rangle~,
\end{align}
where in the first line, we have used Eq. \eqref{3g} to convert
$\zeta_{{\bf k}_2}$ into $\zeta_{g}$ for the first term, and
$\zeta_{{\bf k}_3}$ into $\zeta_{g}$ for the $({\bf
k}_2\Leftrightarrow {\bf k}_3)$ term. The above two terms are
proportional to $k_1^{-3}k_3^{-3}$ and $k_1^{-3}k_2^{-3}$
respectively. We have neglected the term which inserts Eq.
\eqref{3g} for $\zeta_{{\bf k}_1}$, because this term is
proportional to $k_2^{-3}k_3^{-3}$, which can be neglected in the
$k_1\rightarrow 0$ limit.


On the other hand, keeping in mind that a Gaussian random mode of
the field is not affected by other modes, the derivative of
$\zeta_{{\bf k}_2}\zeta_{{\bf
  k}_3}$ takes
the form
\begin{align}\label{redef}
\frac{d}{d\zeta_{{\bf k}_1}^*}(\zeta_{{\bf k}_2}\zeta_{{\bf
  k}_3})&\xrightarrow{k_1\rightarrow 0} \frac{3}{5}f_{\rm NL} \frac{1}{V}
  \sum_{{\bf k}'}
  \frac{d}{d\zeta_{g{\bf k}_1}^*}(\zeta_{g{\bf k}'}\zeta_{g{\bf k}_2-{\bf k}'}\zeta_{g{\bf
  k}_3})+({\bf k}_2\Leftrightarrow {\bf k}_3)\nonumber\\&=
   \frac{12}{5}f_{\rm NL}\frac{1}{V} \zeta_{g{\bf k}_2}\zeta_{g{\bf
  k}_3}~,
\end{align}
where the derivative is taken directly on the operators. So we have
\begin{equation}\label{redef}
\langle\zeta_{{\bf k}_1}\zeta_{{\bf k}_2}\zeta_{{\bf
  k}_3}\rangle\xrightarrow{k_1\rightarrow 0}\frac{1}{V} \langle\zeta_{{\bf k}_1}\zeta_{g{\bf k}_1}^*\rangle
\left\langle\frac{d}{d\zeta_{g{\bf k}_1}^*}(\zeta_{{\bf
k}_2}\zeta_{{\bf
  k}_3})\right\rangle~.
\end{equation}

Generally, when the field before the field redefinition is also
non-Gaussian, one can combine the result in the pervious subsection
and this subsection to obtain the full consistency relation. To be
explicit, if $\delta \phi$ is the quantity in the Hamiltonian, and
$\widetilde{\delta\phi}$ is the quantity we want to calculate in the
correlation function, related by
\begin{equation}
  \widetilde{\delta\phi}_a({\bf x},t)=\sum_b g_{ab}\delta\phi_b({\bf x},t) +
  \sum_{bc} f_{abc}\delta\phi_b({\bf x},t)\delta\phi_c({\bf x},t)~,
\end{equation} then we have
\begin{align}\label{red}
&\langle\widetilde{\delta\phi}_a({\bf
k}_1,t)\widetilde{\delta\phi}_b({\bf
k}_2,t)\widetilde{\delta\phi}_c({\bf
  k}_3,t)\rangle
  \xrightarrow{k_1\rightarrow 0}
  \sum_e \frac{\delta(a,e)}{V}
  \langle\widetilde{\delta\phi}_a({\bf k}_1,t)\delta\phi_e^*({\bf
  k}_1,t)\rangle\nonumber\\&\times \left\{\sum_{fg}g_{bf}g_{cg}
  \frac{\partial}{\partial \delta\phi_e^{{\rm (cl)}*}({\bf
  k}_1,t)}\langle{\delta\phi}_f({\bf
k}_2,t){\delta\phi}_g({\bf
  k}_3,t)\rangle +  \left\langle\frac{\partial}{\partial \delta\phi_e^*({\bf
  k}_1,t)}\left(\widetilde{\delta\phi}_b({\bf
k}_2,t)\widetilde{\delta\phi_c}({\bf
  k}_3,t)\right)\right\rangle\right\}.
\end{align}
Note that in the second line of Eq. \eqref{red}, the derivative in
the first term corresponds to a variation in the Hamiltonian. While
in the second term, the derivative acts as an operator derivative
directly on the field operators $\widetilde{\delta\phi}_b({\bf
k}_2,t)\widetilde{\delta\phi_c}({\bf
  k}_3,t)$. Loosely speaking, we can combine these two derivatives
to write this formula in a more compact form. However, as the
meaning of the two derivatives are not the same, we shall take this
more explicit form as our final result of this subsection.
\subsection{Application to Single Field Inflation}

In this subsection, we shall write the consistency relation
\eqref{3point} in an explicit form in the context of single field
inflation. We shall derive a different expression for the
consistency relation from Maldacena's relation. The new consistency
relation is expressed in terms of the scalar field perturbation in
the flat slice, and it is more convenient to be generalized to the
multi-field case. The derivation for Maldacena's consistency
relation \cite{NG} is given in the appendix. The Maldacena's
relation can also be derived from our consistency relation and a
field redefinition as discussed in the previous subsection (see
Appendix B).

In the flat slice, the inflaton field can be written as
\begin{equation}
  \varphi({\bf x},t)=\varphi(t)+Q({\bf x},t)~,
\end{equation}
where $Q$ can be made gauge invariant in a general slice, known as
the Mukhanov-Sasaki variable \cite{ms}.

 In this case, Eq. \eqref{3point} takes the form
\begin{equation}\label{3pointsingle}
  \langle Q_{{\bf k}_1} Q_{{\bf k}_2} Q_{{\bf k}_3}\rangle
  \xrightarrow{k_1\rightarrow 0} \frac{1}{V}\langle Q_{{\bf k}_1}Q_{{\bf
  k}_1}^*\rangle \frac{d}{d \varphi}\langle Q_{{\bf k}_2} Q_{{\bf
  k}_3}\rangle~.
\end{equation}
This relation is different from Maldacena's consistency relation,
because the transformation from $Q$ to $\zeta$ is nonlinear, and the
shapes of $Q$ and $\zeta$'s 3-point correlation functions are
different.

Eq. \eqref{3pointsingle} can be checked explicitly. For example, in
the single field inflation model with standard kinetic term, the
3-point function in the $k\rightarrow 0$ limit takes the form
\begin{equation}\label{eg1}
  \langle Q_{{\bf k}_1} Q_{{\bf k}_2} Q_{{\bf k}_3}\rangle\rightarrow
  -\frac{1}{V}\frac{\dot\varphi}{H}\langle Q_{{\bf k}_1}Q_{{\bf
  k}_1}^*\rangle \langle Q_{{\bf k}_2} Q_{{\bf k}_3}\rangle~.
\end{equation}
On the other hand, the 2-point function takes the form
\begin{equation}\label{eg2}
  \langle Q_{{\bf k}_2} Q_{{\bf k}_3} \rangle=V \frac{H^2}{2k_2^3}\delta_{{\bf k}_2,-{\bf
  k}_3}~.
\end{equation}
And from $\dot H=-\frac{1}{2}\dot\varphi^2$, the derivative can be
written as \footnote{Strictly speaking, the variation should also
include one more term proportional to $\dot Q$. It is because the
variation of the background we consider is not equal to a time
variation $t\rightarrow t+\delta t$. The correction can be
calculated using $\dot\zeta=0$, so that $\dot
Q=-Q\partial_t(H/\dot\varphi)/(H/\dot\varphi)$. So we should also
set $\dot\varphi\rightarrow \dot\varphi+\dot Q$. However, this
correction is of higher order in the slow roll approximation. So we
neglect this correction in this paper.}
\begin{equation}\label{eg3}
  \frac{d}{dQ}=\frac{d}{d\varphi}=-\frac{1}{2}\dot\varphi
  \frac{d}{dH}~.
\end{equation}
Combining \eqref{eg1}, \eqref{eg2} and \eqref{eg3}, we obtain the
desired relation \eqref{3pointsingle}.

As a more nontrivial test, let us consider the case of generalized
kinetic term. We use the result \cite{Chen:2006nt} for the 3-point
function. To use this result, we need first to know how $\zeta$
corresponds to $Q$ beyond the first order perturbation theory. The
correspondence between $\zeta$ and $Q$ follows directly from the
gauge transformations, so it is not changed after the kinetic term
is generalized. Maldacena's correspondence \cite{NG} between $\zeta$
and $Q$ still holds,
\begin{equation}\label{zetan}
  \zeta=\zeta_n+\frac{1}{2}\left(\frac{\ddot\varphi}{H\dot\varphi}-\frac{\dot H}{H^2}\right)\zeta_n^2+{\rm derivatives}~,\qquad
  \zeta_n\equiv -\frac{HQ}{\dot\varphi}~,
\end{equation}
where the derivatives can be neglected outside the horizon.

We would like to make two remarks at this point. Firstly, In
Maldacena's paper \cite{NG}, equation of motion has been used for
the $\dot H$ term, so the relation in Maldacena's paper only applies
for the standard kinetic term. Here, we do not use any equations of
motion in Eq. \eqref{zetan}. All the terms follow directly from the
gauge transformation. So Eq. \eqref{zetan} holds very generally.
Secondly, the notation $\zeta_n$ is also used in \cite{Chen:2006nt}.
However, the $\zeta_n$ in \cite{Chen:2006nt} is a different
notation, having nothing to do with the relation between $\zeta$ and
$Q$.

In the context of generalized kinetic term, using the equation
$2\dot H=-\dot\varphi^2 P_X$, Eq. \eqref{zetan} can be written as
\begin{equation}
 \zeta=\zeta_n+\frac{1}{4}\left(\eta-\frac{\dot P_X}{HP_X}\right)\zeta_n^2~,\qquad
 \eta\equiv\frac{\dot\epsilon}{H\epsilon}~,\qquad \epsilon\equiv -\frac{\dot
 H}{H^2}~.
\end{equation}
With the above equation in mind, the 3-point correlation function in
\cite{Chen:2006nt} takes the form
\begin{equation}\label{eg4}
  \langle Q_{{\bf k}_1} Q_{{\bf k}_2} Q_{{\bf
  k}_3}\rangle\xrightarrow{k_1\rightarrow 0}
  -\frac{1}{V}\frac{H}{\dot\varphi}\left(2\epsilon+s+\frac{\dot P_X}{HP_X}\right)\langle Q_{{\bf k}_1}Q_{{\bf
  k}_1}^*\rangle \langle Q_{{\bf k}_2} Q_{{\bf k}_3}\rangle~,
\end{equation}
where $s\equiv \dot c_s/(Hc_s)$.

On the other hand, the 2-point correlation function is
\begin{equation}\label{eg5}
  \langle Q_{{\bf k}_2} Q_{{\bf k}_3} \rangle=\frac{V}{2k_2^3} \frac{H^2}{ c_s P_X}\delta_{{\bf k}_2,-{\bf
  k}_3}~.
\end{equation}
Keeping in mind that $\partial_\varphi=\dot\varphi^{-1}\partial_t$,
we can check that the consistency relation Eq. \eqref{3pointsingle}
is respected.

\subsection{Application to Multi-Field Inflation}

The consistency relation in the above subsection can be
straightforwardly generalized to the multi-field case. In this case,
the relation takes the form
\begin{equation}\label{3pointmultiple}
  \langle Q^I_{{\bf k}_1} Q^J_{{\bf k}_2} Q^K_{{\bf k}_3}\rangle
  \xrightarrow{k_1\rightarrow 0} \frac{1}{V}\langle Q^I_{{\bf k}_1}Q^{I*}_{{\bf
  k}_1}\rangle \frac{\partial}{\partial \varphi^I}\langle Q^J_{{\bf k}_2} Q^K_{{\bf
  k}_3}\rangle~.
\end{equation}
where index $I$ in the RHS is not summed. We write
$\partial/\partial\varphi^I$ to denote that the derivative is taken
while other $\varphi^J$ $(J\neq I)$ are fixed.

To check this relation in the multi-field inflation models with
standard kinetic terms (so the field space metric is also flat), we
note that the 3-point correlation function in the $k\rightarrow 0$
limit becomes
\begin{equation}\label{eg4}
 \langle Q_{{\bf k}_1}^I Q_{{\bf k}_2}^J Q_{{\bf k}_3}^K\rangle\rightarrow
  -\frac{1}{V}\frac{\dot\varphi^I }{H}\langle Q_{{\bf k}_1}^IQ_{{\bf
  k}_1}^{I*}\rangle \langle Q_{{\bf k}_2}^J Q_{{\bf k}_3}^K\rangle~.
\end{equation}
The 2-point correlation function takes the form
\begin{equation}\label{eg5}
  \langle Q_{{\bf k}_2}^J Q_{{\bf k}_3}^K \rangle=V \frac{H^2}{2k_2^3}\delta_{{\bf k}_2,-{\bf
  k}_3}\delta^{JK}~.
\end{equation}
From $dH=-\frac{1}{2}\sum_I\dot\varphi^Id\varphi^I$, we have
\begin{equation}\label{eg6}
  \frac{\partial}{\partial\varphi^I}=\frac{\partial H}{\partial
  \varphi^I}\frac{d}{dH}
\end{equation}
Combining \eqref{eg4}, \eqref{eg5} and \eqref{eg6}, we find that the
relation \eqref{3pointmultiple} is indeed satisfied.

In the case with generalized kinetic term, the $1/c_s^2$ order
result is available. In this order, our consistency relation is
satisfied trivially. We hope the above consistency relation can be
checked in some future work for a general sound speed.

In multi-field models, $\zeta$ is usually not conserved. To relate
the quantities at horizon exit with observations, the transfer
function method is usually used. We take double field inflation as
an example \cite{astro-ph/0205253}.

In double field inflation, the inflaton fields can be decomposed
into the inflaton direction $\sigma$ and the perpendicular direction
$s$. The perturbation $Q^I$ can be defined as the perturbation
parallel to the inflation direction ($Q_\sigma$), and perpendicular
to the inflation direction ($Q_s$). These two fields can be further
related to the comoving curvature perturbation $\cal R$ and the
entropy perturbation $\cal S$ as \footnote{The generalization to
non-standard kinetic terms can be found in \cite{Arroja:2008yy}. We
also follow the notation in this paper, but consider only the
standard kinetic term case. When considering the modified kinetic
terms, Eqs. \eqref{a} and \eqref{b} are modified, other discussion
still goes through.}
\begin{equation}\label{a}
  {\cal R}=\frac{HQ_\sigma}{\dot\sigma}~,\qquad {\cal
  S}=\frac{HQ_s}{\dot\sigma}~.
\end{equation}
As shown in \cite{astro-ph/0205253}, the entropy perturbation is
sourceless, and the comoving curvature perturbation is conserved if
there is no entropy perturbation. So the comoving curvature
perturbation in the late times can be written as
\begin{equation}\label{b}
  {\cal R}={\cal R}_*+T_{\cal RS}{\cal S}_*+\delta {\cal R}={\cal A}_\sigma Q_{\sigma*}+
  {\cal A}_sQ_{s*}+\delta {\cal R}~,\qquad {\cal A}_\sigma\equiv
  \frac{H}{\dot\sigma}~,\qquad {\cal A}_s\equiv T_{\cal
  RS}\frac{H}{\dot\sigma}~,
\end{equation}
where $T_{\cal RS}$ can be (at least in principle) determined by
experiments. The term $\delta{\cal R}\sim {\cal O}({\cal R}^2)$
denotes the error in the approximation of taking the first order
comoving curvature perturbation as conserved, as well as using the
linear transfer function beyond the linear perturbation theory. Here
the situation is similar to (but more difficult to handel than) the
difference between $\zeta$ and $\zeta_n$ in the single field case.
To the best of our knowledge, this $\delta {\cal R}$ correction is
not noticed in the literature. We shall leave this correction
uncalculated in this paper.

 The 3-point
function of $\cal R$ takes the form
\begin{align}\label{r}
\langle {\cal R}_{{\bf k}_1} {\cal R}_{{\bf k}_2} {\cal R}_{{\bf
  k}_3}\rangle=\langle({\cal A}_\sigma Q_{\sigma{\bf k}_1}+
  {\cal A}_sQ_{s{\bf k}_1})({\cal A}_\sigma Q_{\sigma{\bf k}_2}+
  {\cal A}_sQ_{s{\bf k}_2})({\cal A}_\sigma Q_{\sigma{\bf k}_3}+
  {\cal A}_sQ_{s{\bf k}_3})\rangle_*+f(\delta{\cal R})~,
\end{align}
where the correlation functions of $Q$ are calculated a few e-folds
after the horizon crossing, and $f(\delta {\cal R})$ denotes the
$\delta {\cal R}$ correction.

The consistency relation can be written as
\begin{align}\label{3m}
  &\langle {\cal R}_{{\bf k}_1} {\cal R}_{{\bf k}_2} {\cal R}_{{\bf
  k}_3}\rangle \xrightarrow{k_1\rightarrow 0} f(\delta{\cal R})+V^{-1}\times \\&\left(
  {\cal A}_\sigma\langle Q_{\sigma{\bf k}_1}Q_{\sigma{\bf k}_1}^*
  \rangle_*\frac{\partial}{\partial\sigma}+{\cal A}_s\langle Q_{s{\bf k}_1}Q_{s{\bf k}_1}^*
  \rangle_*\frac{\partial}{\partial s}\right)
  \langle({\cal A}_\sigma Q_{\sigma{\bf k}_2}+
  {\cal A}_sQ_{s{\bf k}_2})({\cal A}_\sigma Q_{\sigma{\bf k}_3}+
  {\cal A}_sQ_{s{\bf k}_3})\rangle_*~.\nonumber
\end{align}
It can be shown with the standard kinetic terms that the 2-point
cross correlation between the curvature and entropy perturbation
just after the horizon crossing is suppressed by one more order of
slow roll parameters \cite{astro-ph/0205253}. In more general cases,
it is also usually assumed that this correlation can be neglected.
And we note that the entropy direction has the property $\dot s=0$.
Considering the above two conditions, Eq. \eqref{3m} can be
simplified to be
\begin{align}\label{3m1}
  &\langle {\cal R}_{{\bf k}_1} {\cal R}_{{\bf k}_2} {\cal R}_{{\bf
  k}_3}\rangle \xrightarrow{k_1\rightarrow 0} V^{-1}
  {\cal A}_\sigma\langle Q_{\sigma{\bf k}_1}Q_{\sigma{\bf k}_1}^*
  \rangle_*\frac{\partial}{\partial\sigma}\Big(
{\cal A}_\sigma^2  \langle Q_{\sigma{\bf k}_2}Q_{\sigma{\bf
k}_3}\rangle_* +{\cal A}_s^2 \langle Q_{s{\bf k}_2}Q_{s{\bf
k}_3}\rangle_*\Big)+f(\delta{\cal R})~.
\end{align}
This relation is also derived in \cite{Allen:2005ye}. The above
relation can be further simplified in some models. When the kinetic
term is standard, the $Q_\sigma$ power spectrum and the $Q_s$ power
spectrum are equal. When the kinetic term is modified and $c_s\ll
1$, the $Q_s$ power spectrum can be neglected. The relation can be
tested by experiments, on condition that the $f(\delta {\cal R})$
term is calculated.

\subsection{Application to the Curvaton Model}

In the curvaton model, the leading order non-Gaussianity comes from
the field redefinition. In the flat slice, $\zeta$ can be expressed
as
\begin{equation}
  \zeta=\frac{1}{3}\frac{\delta\rho_\sigma}{\rho_\sigma}=\frac{2r}{3}\frac{\delta\sigma}{\sigma}
  +\frac{3}{4r}\left(\frac{2r}{3}\frac{\delta\sigma}{\sigma}\right)^2\equiv
  \zeta_g+\frac{3}{4r}\zeta_g^2~,
\end{equation}
where $r=(3\rho_\sigma)/(4\rho_r+3\rho_\sigma)$, which is calculated
when the curvaton decays. So we have $f_{\rm NL}=5/(4r)$.

The discussion from Eq. \eqref{localNG} to Eq. \eqref{redef} can be
directly applied to the curvaton scenario. The consistency relation
\eqref{redef} is satisfied.

\section{Consistency Relation for the General $n$-point Function}

In this section, we generalize the discussion in the pervious
section to $n$-point correlation functions ($n\geq 3$). The
generalization for the leading order ($N=1$ in Eq. \eqref{pert}) is
straightforward, so we shall not repeat it here. We consider in this
section the general case, including $N\geq 2$.

To save some writing, we use $C(n,N,L)$ to denote the contribution
to the $n$-point correlation function with $L$ loops and order $N$
in Eq. \eqref{pert}. We have
\begin{equation}
\langle \delta\phi_{a_1}({\bf k}_1,t)\cdots \delta\phi_{a_n}({\bf
k}_n,t) \rangle=\sum_{N=0}^{\infty}\sum_{L=0}^{\infty}C(n,N,L)~.
\end{equation}

Note that the number of vertices ($N$), the number of initial lines
($I$) and the number of loops ($L$) satisfy $I=L+N-1$. $C(n,N,L)$
takes the form
\begin{align}\label{C}
  C(n,N,L)
  &=\sum_{\rm PART} i^N
  \int_{-\infty}^{t}dt_{N}\int_{-\infty}^{t_N}dt_{N-1}\cdots\int_{-\infty}^{t_2}dt_{1}
  \nonumber\\&
  \times\left\langle \big[H_{n_1}(t_1),\big[H_{n_2}(t_2),\cdots
  \big[H_{n_N}(t_N),\delta\phi_{a_1}({\bf k}_1,t)\cdots \delta\phi_{a_n}({\bf k}_n,t)\big]\cdots\big]\big] \right\rangle~,
\end{align}
where the ``PART'' in the summation denotes all the partitioning
satisfying $n_1+\cdots+n_N=n+2(L+N-1)$.

Eq. \eqref{H3} can be directly generalized to
\begin{equation}\label{Hn}
  H_{n_i} \xrightarrow{k_1\rightarrow 0,~{\rm contraction}}
  H_{n_i-1}(\bar\phi+\overline{\delta\phi})-H_{n_i-1}(\bar\phi)\equiv
  \Delta H_{n_i-1}~,
\end{equation}
Eqs. \eqref{C} and \eqref{Hn} lead to
\begin{equation}\label{Climit}
  C(n,N,L)\xrightarrow{k_1\rightarrow 0,~{\rm contraction}}\Delta
  C(n-1,N,L)~,
\end{equation}
where $\Delta$ acts on the Hamiltonian and satisfies the Leibnitz
law. From Eq. \eqref{Climit}, we obtain the consistency relation
\begin{align}\label{general1}
&\langle\delta\phi_{a_1}({\bf k}_1,t)\cdots\delta\phi_{a_n}({\bf
  k}_3,t)\rangle \xrightarrow{k_1\rightarrow 0}
\nonumber\\
&  \sum_e \frac{\delta({a_1},e)}{V}
  \langle\delta\phi_{a_1}({\bf k}_1,t)\delta\phi_e^{*}({\bf
  k}_1,t)\rangle
  \frac{\partial}{\partial \delta\phi_e^{{\rm (cl)}*}({\bf
  k}_1,t)}\langle\delta\phi_{a_2}({\bf
k}_2,t)\cdots\delta\phi_{a_n}({\bf
  k}_3,t)\rangle~.
\end{align}
This relation is respected order by order and loop by loop in the
perturbation theory. The consistency relation can also be
iteratively used to obtain the limit that several momenta goes to
zero.

Now consider the case that the quantity in the correlation function
(denoted by $\widetilde{\delta\phi}$) is up to a local field
redefinition compared with the quantity in the Hamiltonian (denoted
by $\delta\phi$). Let the field redefinition be
\begin{equation}
  \widetilde{\delta\phi}_a=\sum_{m=1}^{\infty}\sum_{b_1\ldots b_m}f_{ab_1\ldots b_m}
  \delta\phi_{b_1}*\cdots*\delta\phi_{b_m}~,
\end{equation}
where ``*'' denotes the usual product in the position space and
convolution in the momentum space. One can calculate directly the
LHS and RHS of Eq. \eqref{general1}, replacing $\delta\phi$ with
$\widetilde{\delta\phi}$, so that
\begin{align}\label{general}
&\langle\widetilde{\delta\phi}_{a_1}({\bf
k}_1,t)\cdots\widetilde{\delta\phi}_{a_n}({\bf
  k}_3,t)\rangle
\xrightarrow{k_1\rightarrow 0}  \sum_e \frac{\delta({a_1},e)}{V}
  \langle\widetilde{\delta\phi}_{a_1}({\bf k}_1,t)\delta\phi_e^{*}({\bf
  k}_1,t)\rangle
  \nonumber\\&
  \times\Big\{\sum_{m_2\ldots m_n}^{\infty}\sum_{b_{21}\ldots b_{nm_n}}f_{a_2b_{21}\ldots b_{2m_2}}\cdots f_{a_nb_{n1}\ldots b_{nm_n}}
   \frac{\partial}{\partial \delta\phi_e^{{\rm (cl)}*}({\bf
  k}_1,t)}\langle (\delta\phi_{b_{21}}*\cdots*\delta\phi_{b_{2m_2}})({\bf
k}_1,t)\cdots\rangle \nonumber\\&\qquad+
    \left\langle \frac{\partial}{\partial \delta\phi_{e}^{*}({\bf
  k}_1,t)}(\widetilde{\delta\phi}_{a_2}({\bf
k}_2,t)\cdots\widetilde{\delta\phi}_{a_n}({\bf
  k}_3,t))\right\rangle\Big\}
  ~.
\end{align}
The meaning of the derivatives is the same as that in Eq.
\eqref{red}.

\section{Conclusion}

To conclude, in this paper, we have investigated the consistency
relations for non-Gaussianity.

We have proved the consistency relation for $n$-point correlation
functions dynamically from the Hamiltonian. The proof is model
independent, and valid to all orders of perturbation theory and loop
corrections. We have also derived a consistency relation for local
field redefinitions.

As applications, we have applied the consistency relation to single
field inflation, multi-field inflation, and the curvaton scenario.

For single field inflation, we have got a relation in terms of the
field perturbation $Q$. This relation is different from (but
consistent with) Maldacena's relation. Experimentally, this relation
is not as convenient to use as Maldacena's relation. However, it has
two applications: Firstly, it provides one more theoretical tool to
check the calculation. Secondly, this relation can be
straightforwardly generalized into multi-field case. The single
field 3-point consistency relation is checked explicitly at the
first non-trivial order of slow roll parameters for models with
standard and generalized kinetic terms. We expect that the
consistency relation for the higher order correlation functions can
be checked in the future work.

For multi-field inflation, we have also derived a consistency
relation. The relation is checked explicitly both in the standard
kinetic term case, and the small $c_s$ limit in the generalized
kinetic term case. Future calculation beyond the small $c_s$ limit
should also satisfy this condition. We also point out that one
should notice the $f(\delta{\cal R})$ correction in the calculation.

For the curvaton scenario, we have shown that the consistency
relation is also satisfied, following from  a field redefinition.

\section*{Acknowledgments}
This work is supported by grants of CNSF. We thank Xian Gao and Yang
Zhou for discussions.

\section*{Appendix A: Perturbations in a Box}

In this appendix, we list the relations between the usually used
$\delta$-function normalization and the box normalization used in
this paper. The equations in the left column is for the box
normalization, and the right column corresponds to the
$\delta$-function normalization.

\begin{align}
    \delta\phi_a({\bf x},t)=\frac{1}{\sqrt V}\sum_{\bf k} \delta\phi_a({\bf
  k},t) e^{i {\bf k}\cdot {\bf x}} && \Leftrightarrow &&
   \delta\phi_a({\bf x},t)=\int\frac{d^3k}{(2\pi)^3} \delta\phi_a({\bf
  k},t) e^{i {\bf k}\cdot {\bf x}}~,\nonumber\\
  \delta\phi_a({\bf k},t)=\frac{1}{\sqrt V}\int d^3x \delta\phi_a({\bf
  x},t) e^{-i {\bf k}\cdot {\bf x}} && \Leftrightarrow &&
  \delta\phi_a({\bf k},t)=\int d^3x \delta\phi_a({\bf
  x},t) e^{-i {\bf k}\cdot {\bf x}}~,\nonumber\\
  V\delta_{{\bf k},{\bf k}'} && \Leftrightarrow &&
  (2\pi)^3\delta^3({\bf k}-{\bf k}')~,\nonumber\\
  [a_{\bf k}, a^\dagger_{{\bf k}'}]=\delta_{{\bf k},{\bf k}'}
  &&\Leftrightarrow && [a_{\bf k}, a^\dagger_{{\bf
  k}'}]=(2\pi)^3\delta^3({\bf k}-{\bf k}')~.
\end{align}

\section*{Appendix B: Maldacena's Consistency Relation}

In the single field case, the quantity one usually calculates in a
correlation function is the curvature perturbation in the uniform
density slice. This curvature perturbation $\zeta$ can be thought of
as a local rescaling of the scale factor. When calculating the two
point function, the derivative with respect to $\zeta$ can be
translated to derivative with respective to $\tilde a$ as
\begin{equation}
  \frac{1}{\sqrt V}\frac{d}{d\overline\zeta}=\frac{d}{d\zeta_{{\bf k}_1}}=-\frac{d}{d \ln
  a_*}~,
\end{equation}
The minus sign before $d/(d\ln a_*)$ (``$*$'' denotes the time of
horizon crossing) appears because the rescaling of the scale factor
only affects the horizon crossing time. The local Hubble constant is
not affected. So a positive $\zeta_{{\bf k}_1}$ corresponds to a
earlier horizon crossing time.

In this context, Eq. \eqref{3point} becomes
\begin{equation}\label{3m}
  \langle\zeta_{{\bf k}_1}\zeta_{{\bf k}_2}\zeta_{{\bf
  k}_3}\rangle\xrightarrow{k_1\rightarrow 0}-\frac{1}{V}\langle\zeta_{{\bf
  k}_1}\zeta_{{\bf
  k}_1}^*\rangle \frac{d}{d\ln a_*}\langle\zeta_{{\bf k}_2}\zeta_{{\bf
  k}_3}\rangle~.
\end{equation}
This is the well-known consistency relation for single field
inflation.

This consistency relation can also be derived from Eq.
\eqref{3pointsingle} and a field redefinition. Up to second order,
$\zeta$ and $Q$ are related by Eq. \eqref{zetan}. Using Eqs.
\eqref{red} and \eqref{3pointsingle}, we have
\begin{equation}\label{dr}
\langle\zeta_{{\bf k}_1}\zeta_{{\bf k}_2}\zeta_{{\bf
  k}_3}\rangle\xrightarrow{k_1\rightarrow
  0}\frac{1}{V}\left(-\frac{H}{\dot\varphi}\right)\langle\zeta_{{\bf k}_1}
\zeta_{{\bf k}_1}^*\rangle\frac{d}{d Q_{{\bf k}_1}^{{\rm
(cl)}*}}\langle Q_{{\bf k}_2} Q_{{\bf
k}_3}\rangle+\frac{2}{V}\left(\frac{\ddot\varphi}{H\dot\varphi}-\frac{\dot
H}{H^2}\right)\langle\zeta_{{\bf
  k}_1}\zeta_{{\bf
  k}_1}^*\rangle \langle\zeta_{{\bf k}_2}\zeta_{{\bf
  k}_3}\rangle~.
\end{equation}
Using $d/(d Q_{{\bf k}_1}^{{\rm (cl)}*})=d/(d
\varphi)=d/(\dot\varphi d t)$, we find that the two terms in the RHS
of \eqref{dr} can be combined to obtain the consistency relation
\eqref{3m}.

 Similarly, we can recover the consistency relation for
$\langle\zeta_{{\bf k}_1}\gamma^{s_2}_{{\bf k}_2}\gamma^{s_3}_{{\bf
  k}_3}\rangle$, where $\gamma^{s}_{\bf k}$ denotes the perturbation of
  gravitational waves, and $s$ denotes the polarization ($s=1,2$). The
consistency  relation takes the form \cite{NG}
\begin{equation}
   \langle\zeta_{{\bf k}_1}\gamma^{s_2}_{{\bf k}_2}\gamma^{s_3}_{{\bf
  k}_3}\rangle\xrightarrow{k_1\rightarrow 0}-\frac{1}{V}\langle\zeta_{{\bf
  k}_1}\zeta_{{\bf
  k}_1}^*\rangle \frac{d}{d\ln a_*}\langle\gamma^{s_2}_{{\bf k}_2}\gamma^{s_3}_{{\bf
  k}_3}\rangle~.
\end{equation}

To derive the consistency relation for $\langle \gamma^{s_1}_{{\bf
  k}_1}\zeta_{{\bf
k}_2}\zeta_{{\bf k}_3}\rangle$, one need to know how the universe
locally look like
  with the existence of a long wave length gravitational wave. As
  discussed in \cite{NG}, this effect corresponds to
  the change $k^2\rightarrow k^2-\gamma_{ij}k^ik^j$. So the
  consistency relation becomes
\begin{equation}
  \langle\gamma^{s_1}_{{\bf k}_1}\zeta_{{\bf k}_2}\zeta_{{\bf
  k}_3}\rangle\xrightarrow{k_1\rightarrow 0}-\frac{1}{V}\langle\gamma^{s_1}_{{\bf
  k}_1}\gamma^{s_1*}_{{\bf
  k}_1}\rangle\epsilon_{ij}^{s_1}k_2^ik_2^j \frac{d}{dk_2^2}\langle\zeta_{{\bf k}_2}\zeta_{{\bf
  k}_3}\rangle~.
\end{equation}


\end{document}